\documentstyle[11pt]{article}
\input epsf

\begin{document}

\baselineskip 24pt

\title{\Large\bf A strongly first order electroweak phase transition\\
                 from strong symmetry-breaking interactions}
\author{Thomas Appelquist\thanks{Electronic address:
 twa@genesis3.physics.yale.edu},
 Myckola Schwetz\thanks{Electronic address:
 ms@genesis2.physics.yale.edu}\ \ and
 Stephen B. Selipsky\thanks{Electronic address:
 stephen@genesis1.physics.yale.edu}\\ \\ {\it
 Department of Physics, Yale University, New Haven, Connecticut 06520-8120}}

\date{February 23, 1995}

\maketitle

\begin{picture}(0,0)(0,0)
\put(365,235){hep-ph/9502387}
\put(365,220){YCTP-P1-95A}
\end{picture}
\vspace{-24pt}

\begin{abstract}
We argue that a strongly first order electroweak phase transition is natural
in the presence of strong symmetry-breaking interactions, such as technicolor.
We demonstrate this using an effective linear scalar theory of the
symmetry-breaking sector.
\end{abstract}

\section{Introduction} \label{intro:sec}

   The nature of the finite temperature electroweak phase transition remains
an important and elusive problem in particle physics.  The possibility of
electroweak scale baryogenesis, which has helped generate new interest in
this subject \cite{baryogenreview}, requires the transition to be strongly
first order; only then does the universe depart from thermal equilibrium,
satisfying the third of Sakharov's requirements for baryogenesis and
avoiding the washing out of any generated baryon asymmetry.
This paper investigates the electroweak phase transition when the
symmetry-breaking sector involves strong interactions, such as technicolor.
We use an effective scalar theory to describe these interactions,
and find that the transition is first order if this sector exhibits a
$U(N)_L\times U(N)_R$ global symmetry.  When the interactions are relatively
strong, the transition is strongly first order:  the discontinuity in the
order parameter at the critical temperature is of the order of the zero
temperature expectation value $v$.

   To set the stage, we briefly review what is known about the electroweak
phase transition in the minimal standard model.  A variety of studies of
the finite temperature effective potential have addressed this problem,
using methods including the $\epsilon$ expansion, Wilson-effective-action
and ``average action'' renormalization group techniques, or graphical
summation schemes with names such as daisy and superdaisy
\cite{priorSM,ArnoldEspinosa}.  They allow improved treatments of the
infrared problems \cite{WeinbergIR} that can arise near phase transitions.
The transition appears to be first order when the Higgs boson is lighter
than the $W$ and $Z$, but at most rather weakly so for an experimentally
acceptable Higgs boson mass.  As that mass increases, corresponding to stronger
scalar coupling, the strength of the transition decreases \cite{latticeSU2}.
When the Higgs boson is much heavier than the $W$ and $Z$, it is reasonable
to neglect the gauge and Yukawa interactions when studying the transition.
In that approximation the minimal standard model is the $O(4)$ linear sigma
model, and the chiral transition in $O(N)$ models seems to be second order
or at most weakly first order \cite{phi4order2,Jain,Ginsparg3d}.

   Beyond the minimal standard model, less is known about the nature of the
electroweak phase transition, especially if the symmetry-breaking sector is
strongly interacting, as in a technicolor theory.
Lattice studies \cite{latticegauge} of QCD-like gauge theories indicate that
with two light fermions, corresponding to an $SU(2)_L\times SU(2)_R \sim O(4)$
global symmetry, the transition is second order, whereas with more light
fermions and a larger global symmetry, the transition is first order.
This suggests that technicolor theories, which often do have more than two
technifermions, might lead to a first order transition.  To complement the
lattice results, we analyze the strongly coupled symmetry-breaking problem
using an effective scalar theory, which is partly amenable to analytic
treatment and avoids some of the uncertainties associated with lattice
techniques.

\section{The Linear Sigma Model} \label{model:sec}

   The properties of the phase transition depend on the dynamics of the
electroweak order parameter.  We assign it to transform under a global
$U(N)_L\times U(N)_R$ chiral symmetry, motivated by technicolor theories,
in which the order parameter is a condensate of fermion bilinears.
A one family technicolor model for example corresponds to $N=8$ flavors,
large enough to suggest utility of the large $N$ expansion in the analysis.
The most economical effective scalar theory of electroweak interactions is
of course the nonlinear chiral Lagrangian.
To allow for the existence of a phase transition above which the order
parameter vanishes, however, we must go beyond the nonlinear theory
\cite{NonlinearChiralPhase}.  Accordingly, we will use an effective linear
sigma model to describe the symmetry-breaking sector
\cite{PisarskiWilczek,otherlinear}.  Neglecting standard model gauge and
fermion interactions in comparison to this sector's strong interactions,
we take the Lagrangian to be
\begin{equation} \label{sigmalagrangian}
 {\cal L} = {\rm Tr}\left( \partial_{\mu}\Sigma^\dagger \partial^\mu \Sigma
    \right)
  - \mu^2 {\rm Tr}\left( \Sigma^\dagger \Sigma \right)
  - {\lambda_1\over{4N^2}}\left({\rm Tr}\, \Sigma^\dagger \Sigma \right)^2
  - {\lambda_2 \over{4N}}{\rm Tr}\, {\cal O}_2^2 \ .
\end{equation}
Here we have defined the traceless operator
${\cal O}_2 \equiv \Sigma^\dagger \Sigma - ({\cal I}/N)
 \ {\rm Tr}\, (\Sigma^\dagger \Sigma)$, where
\begin{equation} \label{sigmacomponent}
  \Sigma = (\sigma_a + i\pi_a)\ T^a \qquad (a = 0,1,\ldots , N^2 - 1)\ ,
\end{equation}
with generator matrix normalization $2\ {\rm Tr}\, (T^a T^b) = \delta^{ab}$
and $T^0 = {\cal I}/\sqrt{2N}$.  Higher-dimensional (nonrenormalizable)
interactions could also be included in the effective theory, but we
argue in section~\ref{conclusions:sec} that their omission does not
qualitatively affect our conclusions.

   In order that the tree level potential be bounded below we must have
$\lambda_1 > 0$, and either $\lambda_2 > 0$, or $\lambda_2 < 0$ with
$\lambda_1 > (N-1) |\lambda_2|$.  Spontaneous symmetry-breaking then occurs
for negative $\mu^2$.  We do not consider the $\lambda_2 < 0$ case, in which
$U(N)_L\times U(N)_R$ breaks to
$[U(N-1)\times U(1)]_L \times [U(N-1)\times U(1)]_R$.  For $\lambda_2 > 0$
the breaking pattern is $U(N)_L\times U(N)_R \rightarrow U(N)_V$, and the
tree level potential at zero temperature is minimized by a vacuum expectation
value $\langle\Omega|\Sigma|\Omega\rangle$ that can be taken, after a
$U(N)_L\times U(N)_R$ transformation, to be real and proportional to
the identity matrix:
\begin{eqnarray} \label{treevev}
  &\langle\Omega|\sigma_0|\Omega\rangle^2 =& {4N^2\over\lambda_1} |\mu^2|
 \ ,\nonumber\\
  &\langle\Omega|\sigma_a|\Omega\rangle =& 0\qquad (a = 1,\ldots , N^2 -1)
 \ ,\nonumber\\
  &\langle\Omega|\pi_a|\Omega\rangle =& 0\qquad (a = 0,\ldots , N^2 -1) \ .
\end{eqnarray}
The $\lambda_2$ independence of $\langle\Omega|\sigma_0|\Omega\rangle$
follows from the definition of ${\cal O}_2$.  The $W$ mass is given by
${1\over 2} g \langle\sigma_0\rangle$, so that
$\langle\Omega|\sigma_0|\Omega\rangle\ = v \equiv 250$ GeV.
The $2N^2$ degrees of freedom described by the $\Sigma$ fields correspond
to $N^2$ Nambu--Goldstone scalars $\pi_a$, and $N^2$ scalars ($\sigma_a,
\sigma_0$) that describe the massive fluctuations of the order parameter.
Their masses
\begin{equation} \label{vacmasses}
  m^2_{\pi_a}= 0\ ,\ \
  m^2_{\sigma_a} = {\lambda_2 \over {2N^2}} v^2\ ,\ \
  m^2_{\sigma_0} = {{\lambda_1 + \lambda_2} \over {2N^2}} v^2
\end{equation}
scale as $1/N$ relative to $v = 250$ GeV.

   Among the $N^2$ Nambu--Goldstone scalars, three become the longitudinal
$W^\pm$ and $Z$, the $\pi_0$ gains mass via anomalous breaking of the axial
$U(1)$, and the remaining $(N^2 -4)$ are pseudo-Nambu--Goldstone bosons,
all of which gain small masses from neglected standard model and other
(Extended Technicolor) interactions.
We neglect these pseudo-Nambu--Goldstone masses.
The anomaly-generated $\pi_0$ mass could also be implemented in our effective
Lagrangian, by adding a nonrenormalizable (for $N > 4$) determinantal
interaction \cite{PisarskiWilczek,otherlinear,DetTerm}, but omission of
this term affects the masslessness of only one mode and will not change
our conclusions.  
In the limit $\lambda_2 = 0$, the symmetry of the theory increases from
$U(N)\times U(N)$ to $O(2N^2)$ and the spontaneous breaking produces
$(2N^2 -1)$ Nambu--Goldstone bosons, with only the $\sigma_0$ massive.

   The factors of $N$ in Eq.~(\ref{sigmalagrangian}) allow a nontrivial
large $N$ expansion holding $\lambda_1$ and $\lambda_2$ fixed.
In the $\lambda_2 = 0$ limit, the leading $1/N$ approximation is tractable,
corresponding to the familiar linear bubble sum.  With $\lambda_2$ nonzero,
on the other hand, all planar diagrams involving the $\lambda_2$ interaction
contribute at leading order.  For both interactions, the strong coupling
limit sets in at $\lambda_i /16 \pi^2 \sim 1$, when higher loops are as
large as lower order contributions.  In this limit the scalar masses of
Eq.~(\ref{vacmasses}) saturate the bound $4\pi v/N$ \cite{SoldateSundrum}.

   Symmetry-breaking aspects of the $U(N)\times U(N)$ linear sigma model at
zero temperature in various spatial dimensions have previously been examined
using the $\epsilon$ expansion \cite{PisarskiWilczek} combined with the
effective potential \cite{Paterson}, and with lattice methods \cite{YueShen}.
These investigations found Coleman--Weinberg \cite{CW} behavior:
broken symmetry when $\mu^2$ is tuned to zero.  The $\epsilon$ expansion
analysis, using renormalization group flow in $d = 4 - \epsilon$ spatial
dimensions, shows that a second order transition is not self-consistent for
$N > 2$ \cite{PisarskiWilczek}.  However, studying the transition as a
function of Lagrangian parameters
at $T = 0$ is not the same as studying it as a function of temperature.
Since effective three-dimensional theories apply to finite temperature only
by assuming decoupling of nonstatic modes (discussed below), they are limited
to the second order or very weakly first order cases.  The same is true of
analyses of the transition strength using renormalization group flow ``time''
\cite{RGflowstrength}.  Furthermore, the $\epsilon$ expansion may break down
before $\epsilon = 1$ or miss IR fixed points.  Nevertheless, the above work
does hint that a full finite temperature calculation might find a first order
phase transition.

\section{Computing the Effective Potential} \label{computing:sec}

   We analyze the phase transition with the finite temperature effective
potential defined in Euclidean periodic time, which describes the system
in thermal equilibrium \cite{DolanJackiw}.  Matsubara frequencies
$k^0 = (2\pi n\, T)$, where $n = 0,\pm 1, \pm 2,\ldots$, then appear in
propagators.  If the relevant momentum scales were much smaller than
$2\pi\, T$ (e.g.\ the high temperature limit), all but the $n = 0$
modes would decouple, leaving an effective three-dimensional theory.
This is not true at a strongly first order phase transition,
where we shall see that relevant scales are of order $2\pi\, T$.
(The transition temperature will be of order $v/N$, while in the strongly
coupled case the masses in Eq.~(\ref{vacmasses}) approach $4\pi\, v/N$.)
We will therefore retain the full four-dimensional dynamics; we will also
avoid the relative simplicity of the high temperature expansion, relying
below on numerical integration for explicit results.

   We can obtain the finite temperature effective potential as the sum of
one-particle-irreducible vacuum graphs in a background field \cite{Jackiw}.
For the linear sigma model of interest here, the calculation is conveniently
implemented using the auxiliary field method \cite{auxfield,CarenaWagner}
to eliminate the $\lambda_1$ interaction in favor of a nonpropagating
dimension-two field $\chi$.  The auxiliary field facilitates resummation
to all orders in $\lambda_1$, at each order in $1/N$.
It consists of adding to the Lagrangian a perfect square, which yields an
irrelevant constant factor upon path integration over the auxiliary field:
\begin{equation} \label{auxlagrangian}
  {\cal L} \rightarrow {\cal L} + {4N^2\over \lambda_1}
   \left(\chi - \mu^2 - {\lambda_1\over {4N^2}}\,
   {\rm Tr}\, (\Sigma^\dagger \Sigma)\right)^2 \ .
\end{equation}

   Before presenting results for the full theory, it is useful to review
the nature of the phase transition in the $\lambda_2 = 0$ theory ($O(2N^2)$
symmetry).  Analyses using the $1/N$ expansion show that the high temperature
phase transition here is second order \cite{Jain,CarenaWagner,TetradisLargeN}.
The leading order computation of the effective potential corresponds to the
``superdaisy'' approximation \cite{DolanJackiw,Fendley,BBHsu}, which can
therefore be justified in the context of the $1/N$ expansion.  Without
superdaisy resummation, on the other hand, a one-loop computation using
the Lagrangian of Eq.~(\ref{sigmalagrangian}) with $\lambda_2 = 0$ gives
a first order result \cite{ArnoldEspinosa,EQZ}.  This is unreliable,
since the theory's infrared behavior is controlled by the effective
loop expansion parameter, of order $(\lambda_1/16\pi^2)\ T/m(\sigma)$.
Here $m^2(\sigma) \equiv \mu^2 + (\lambda_1/4) \sigma^2$ is the effective
mass of the lightest excitations, in the quantum state $|\Psi\rangle$ with
amplitude peaked at the classical background field
$\langle\Psi|\sigma_0|\Psi\rangle \equiv N\sigma$.  This mass vanishes for
Nambu--Goldstone bosons at a symmetry-breaking minimum of the effective
potential, so the perturbative expansion breaks down due to infrared
divergences, in some $\sigma$ neighborhood of the minimum.  The $1/N$
expansion avoids this perturbative problem by a summation of graphs at
each order in $1/N$.  

   To leading order in $1/N$, the $O(2N^2)$ computation involves only a single
loop of the $2N^2 - 1$ Nambu--Goldstone bosons in the presence of background
$\chi$ and $\sigma$ fields, using the Lagrangian of Eq.~(\ref{auxlagrangian}).
Because the $\chi$ field is nondynamical, it can be eliminated in favor of
$\sigma$ using the saddle-point equation of motion.  This procedure reveals
the second order behavior.  It is equivalent to using the Lagrangian of
Eq.~(\ref{sigmalagrangian}) and solving a momentum-independent Schwinger--Dyson
equation for the Nambu--Goldstone boson effective mass, whose solution is
inserted into the single loop of Nambu--Goldstone bosons.

   The effective potential computed in this way becomes complex for $\sigma$
between the symmetry-breaking minimum (where the Nambu--Goldstone boson mass
vanishes) and the origin.  This occurs for $\lambda_2 \neq 0$ as well;
in either case, the imaginary part may be interpreted \cite{WeinWu} as
the transition amplitude for leaving an unstable configuration, into which
a classical external source cannot force the system.  One may define a real
quantity by a Maxwell construction:  the ``convex effective potential''
(the convex envelope of the real part of the above effective potential)
\cite{convex}.  Although the convex effective potential approximately
describes the lowest energy quantum state for a given $\sigma$, if a local
minimum remains at the origin then the system is likely to supercool,
remaining in the quantum state peaked at $\sigma = 0$ even below the
temperature $T_c$ where the symmetry-breaking minimum falls to a lower
energy than the symmetric one \cite{falsevacuum}.  The phase transition
is thus controlled by the appearance and location of minima in the nonconvex
potential as computed here.

\begin{figure}[hbt]
\hfil{
 \epsfysize = 0.8in
 \epsfbox{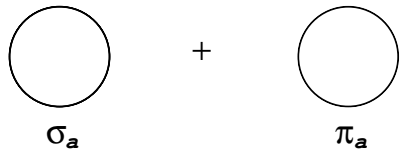}
 }\hfil
\caption{Feynman graphs included in our computation of $V_{\rm eff}$.
 The vacuum loops of $(N^2)$ flavors of $\pi_a$ and $(N^2 -1)$ of
 $\sigma_a$ propagate in background $\sigma_0$ and $\chi$ fields.}
\label{graphsin}
\end{figure}

\begin{figure}[hbt]
\hfil{
 \epsfxsize = 5in
 \epsfbox{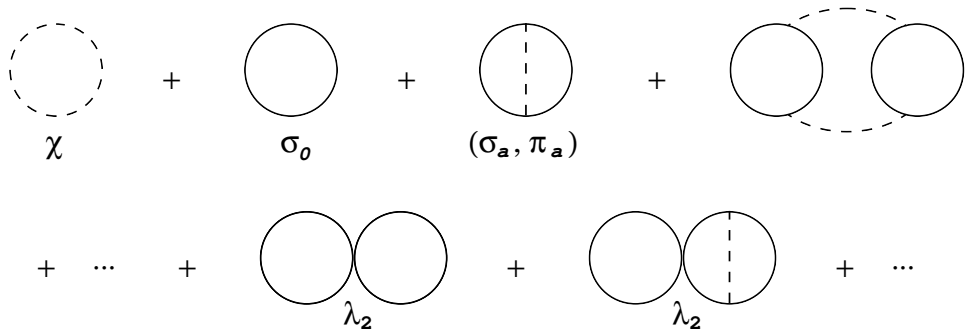}
 }\hfil
\caption{Some neglected graphs.  One-loop graphs subleading in $1/N$; and
 higher order graphs suppressed by powers of $1/N$ or $\lambda_2$ or both.
 Dashed lines are diagonalized $\chi$ field propagators, solid lines are
 diagonalized $\sigma$ and $\pi$ fields.}
\label{graphsout}
\end{figure}

   Turning now to the general case with $\lambda_2 \neq 0$, the computation
is complicated by the fact that planar graphs to all orders in $\lambda_2$
contribute at leading order in the $1/N$ expansion.  Auxiliary fields could
also be introduced to replace the $\lambda_2$ interaction, but not very
usefully since an infinite class of graphs would still contribute at each
order in $1/N$.  We instead use the Lagrangian with only the auxiliary
$\chi$ field, Eq.~(\ref{auxlagrangian}), and compute the effective potential
as we did in the $\lambda_2 = 0$ case:  one loop of $(2N^2 - 1)$ quanta in
background $\chi$ and $\sigma$ fields (Fig.~\ref{graphsin}), after which the
auxiliary $\chi$ field is eliminated using its equation of motion.
This approximation sums all orders in $\lambda_1$ to leading order in $1/N$.
It also includes the (tree-level) contribution of the $\lambda_2$ interaction
to the effective masses of the $(2N^2 -1)$ quanta in the loop, lifting the
degeneracy between the $N^2$ Nambu--Goldstone $\pi_a$ modes and the $(N^2 - 1)$
$\sigma_a$ modes which are Nambu--Goldstone bosons only in the
$\lambda_2\rightarrow 0$ limit.
This approximation to the $U(N)\times U(N)$ theory should provide a reliable
qualitative guide to the nature of the phase transition, even without
contributions like those of Fig.~\ref{graphsout}.

   The computation requires renormalization counterterms which we define
conventionally at zero temperature.  Diagonalizing the propagators then
gives the one-loop result for the finite temperature effective potential
as a function of $\sigma \equiv\ \langle\Psi|\sigma_0/N|\Psi\rangle$ and
$\chi \equiv \langle\Psi|\chi|\Psi\rangle$:
\begin{eqnarray} \label{logdet}
 N^{-2}V(\sigma,\chi) &=& {1\over 2}\chi\sigma^2 +
  2{\mu^2\chi\over\lambda_1} - {\chi^2\over\lambda_1}
 \nonumber\\ &+&\ {1\over 2}\int_T{\ln{(k^2+\chi)}} +\
  {1\over 2}\int_T{\ln{(k^2+\chi+{\lambda_2\over 2}\sigma^2)}}
 \nonumber\\ &-&\ (T=0\ {\rm counterterms})
\end{eqnarray}
where $k^0 = (2\pi n T)$ and
$\int_T \equiv T\int (2\pi)^{-3}d^3k \sum_{n=-\infty}^{n=\infty}$.
The first integral corresponds to the loop of $\pi_a$ (Nambu--Goldstone)
scalars whose effective mass is $\chi$; the second integral corresponds
to the loop of $\sigma_a$ scalars whose effective mass is
$\chi+(\lambda_2/2)\sigma^2$.  The subtracted counterterms are
\begin{eqnarray} \label{counterterms}
  \int_0 \ln k^2 + \int_0
  {{\chi+{\lambda_2\over 2N^2}{\rm Tr}\,(\Sigma^{\dagger}\Sigma)}\over k^2}
  - {1\over 4}\int_0 {{\chi^2 + \left[
   \chi+{\lambda_2\over N^2}{\rm Tr}\,(\Sigma^{\dagger}\Sigma)\right]^2}
  \over (k^2 + M_r^2)^2} \nonumber\\
 =\ \int_0 \ln k^2 \ +\ \int_0 {{\chi+{\lambda_2\over 4}\sigma^2}\over k^2}
 \ -\ {1\over 4}\int_0
    {{\chi^2 + (\chi+{\lambda_2\over 2}\sigma^2)^2}\over (k^2 + M_r^2)^2}\ ,
\end{eqnarray}
where the subscript $0$ on the integrals refers to zero temperature.
The first term is an unobservable vacuum energy and the second corresponds
to renormalization of $\mu^2$.  In the final term, which corresponds to
$\mu^2$ and $\lambda_1$ renormalization, we have inserted a renormalization
scale in the form of an infrared cutoff on the integral.  Physical results
are not sensitive to details of the renormalization scheme.

   The renormalization of $\lambda_1$ to leading order in $1/N$ includes all
$\lambda_1$-dependent corrections, and a subset of the $\lambda_2$-dependent
corrections.  There is no $\lambda_2$ renormalization in our approximation
(with a purely $\sigma_0$ background).  The corrections to $\lambda_1$ lead
to a Landau pole at a sufficiently high scale, reflected in a corresponding
pathology in the effective potential at a sufficiently high value of $\sigma$.
We avoid this problem by taking $\lambda_1$ to be somewhat less than the
strong coupling limit at the scale $M_r$, which in turn we take to lie
just above the momentum scales relevant in the effective potential.
Thus the Landau pole is safely out of range at the scales of interest.

   Carrying out the summation in Eq.~(\ref{logdet}) gives
\begin{eqnarray} \label{effpotential}
 N^{-2}V(\sigma,\chi) &=& {1\over 2}\chi\sigma^2
  + {2\mu^2\chi\over\lambda_1} - {\chi^2\over\lambda_1} +
  {\chi^2\over{64\pi^2}}\left( \ln{\chi\over M_r^2} - {3\over 2} \right)
\nonumber\\ &+& {T^4\over{2\pi^2}} \int_0^\infty x^2 dx
 \ln{\left(1 - \exp{\left[-\sqrt{x^2 + \chi/T^2}\right]}\right)}
\nonumber\\ &+&
 {(\chi+{\lambda_2\sigma^2\over 2})^2\over{64\pi^2}}
 \left(\ln{(\chi+{\lambda_2\sigma^2\over 2})\over M_r^2}-{3\over 2}\right)
\nonumber\\ &+& {T^4\over{2\pi^2}} \int_0^\infty x^2 dx
 \ln{\left(1 - \exp{\left[-\sqrt{x^2 +
 { {(\chi+{\lambda_2\sigma^2\over 2})}\over{T^2} }}\right]}\right)} .
\end{eqnarray}
The auxiliary field is eliminated by solving its equation of motion
\begin{equation} \label{dvdchi}
 \left({\partial V(\sigma,\chi)\over\partial\chi}\right)_\sigma = 0\ .
\end{equation}
This transcendental equation, which we solve numerically, is equivalent
to a single Schwinger--Dyson equation for the scalar masses.  Had we
introduced another auxiliary field to resum the $\lambda_2$ interaction,
we would have coupled Schwinger--Dyson equations for the scalar masses.
Substituting the solution $\chi(\sigma)$ into $V(\sigma,\chi)$ gives
the finite temperature effective potential as a function of $\sigma$ alone:
$V(\chi(\sigma),\sigma) \equiv U(\sigma)$.  We have carried out numerical
computations of $U(\sigma)$ for a range of coupling strengths and temperatures.

\begin{figure}[hbt]
\hfil{
 \epsfysize = 3in
 \epsfbox{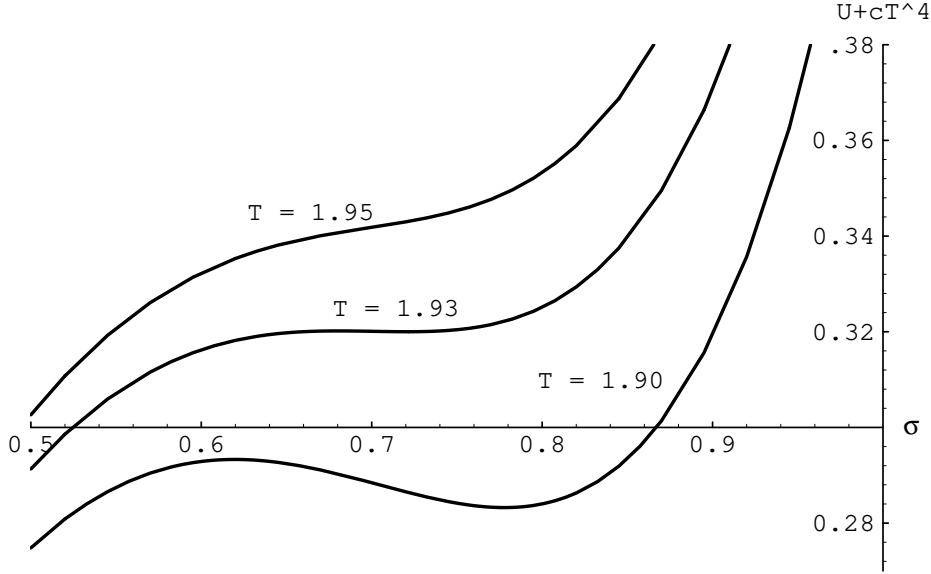}
 }\hfil
\caption{The effective potential $N^{-2}U(\sigma)$ for several temperatures,
 shifted by $(\pi^2/45)\ T^4$ for clarity,  with all dimensionful quantities
 measured in units of $v/N$.  A first order phase transition must occur below
 the temperature $T_{\rm inflect} = 1.93 v/N$ where the potential
 maximum and minimum coincide.  The field strength discontinuity is at least
 $0.7\ v/N$ here, with parameters $\lambda_1(M_r = 4.54 v/N)= 40 =\lambda_2$.}
\label{plotU}
\end{figure}

   In Fig.~\ref{plotU} we present numerical results for the choice of couplings
$\lambda_1(M_r)/16\pi^2 = \lambda_2/16\pi^2 \approx 0.25$, where we have taken
$M_r \approx 4.5 v/N$.  This value for the loop expansion parameters is large,
but still in a range that keeps the scalar masses, Eq.~(\ref{vacmasses}),
somewhat below $4\pi v/N$, and the Landau pole above the momentum range
relevant to the problem.  For various temperatures, we plot
$N^{-2} U(\sigma) + (\pi^2/45)\ T^4$
for a $\sigma$ range over which $U$ is real.
The purely temperature-dependent second term merely shifts the curves onto
the same scale, preserving their shapes and ordering.

   For $\sigma$ values somewhat smaller than those shown in the curves,
below where the Nambu--Goldstone propagator mass $\chi$ vanishes,
$U(\sigma)$ develops a complex part just as in the $O(2N^2)$ case.
However, with $\lambda_2 \neq 0$ the propagator masses in our approximation
do not include all terms of higher order in $\lambda_2$ at leading order in
$1/N$.  Goldstone's theorem, applied at a minimum of the {\it effective}
potential, refers not to such propagator masses calculated at lower order,
but instead to second derivatives in symmetry directions.  Thus $\chi$ can
vanish closer to the origin than the symmetry-breaking minimum.  With the
parameters of Fig.~\ref{plotU}, $T/{\sqrt\chi}$ is close to unity over the
$\sigma$ range shown; we thus avoid infrared divergences at the minimum.

   The strongly first order character of the phase transition is apparent
 from the curves, in which the symmetry-breaking local minimum first
appears at a field value $\sigma_{\rm min}\approx 0.7 v/N$, at temperature
$T_{\rm inflect}$ just below $2 v/N$.  With decreasing temperature
the minimum evolves to larger field values until it becomes the usual $T=0$
symmetry-breaking vacuum with $\sigma_{\rm min} = v/N$.  We have checked
that as $\lambda_2$ decreases, $\sigma_{\rm min}(T_{\rm inflect})$
also decreases, with $T_{\rm inflect}$ varying only slightly.
This behavior is consistent with the second order nature of the phase
transition in the $\lambda_2 = 0$ limit.

   We can gain some qualitative insight into the origin of first order
behavior in the $\lambda_2 \neq 0$ theory by contrasting it with the
$\lambda_2 = 0$ theory.  There, all but one of the degrees of freedom are
Nambu--Goldstone modes, whose mass $\chi$ vanishes at the potential minimum
$\sigma_{\rm min}$.  The vanishing of $\chi$ leaves Eq.~(\ref{dvdchi}) at
$\sigma_{\rm min}$ in the simple form
$\sigma_{\rm min}^2 - (v/N)^2 + T^2/6 = 0$, clearly showing second order
behavior at $T = T_c \equiv {\sqrt 6} v/N$.  The relatively small masses
generated by standard model gauge interactions would in fact turn these
modes into pseudo-Nambu--Goldstone bosons, but such small perturbations
induce at most weakly first order behavior.  Even without those interactions,
for small nonzero $\lambda_2$ the $(N^2 - 1)\ \sigma_a$ modes may be thought
of as pseudo-Nambu--Goldstone bosons of the $O(2N^2)$ theory, and their
masses do not vanish with the $\pi_a$ mass.  For large $\lambda_2$, however,
this is a large perturbation, preventing the equation for $\sigma_{\rm min}$
from taking on the simple second order form above.  The full transcendental
equation in our case leads to strongly first order behavior.

\section{Conclusions} \label{conclusions:sec}

   We have argued using an effective linear sigma model that the high
temperature electroweak phase transition, in the presence of a strongly
interacting symmetry-breaking sector, is strongly first order.
With the global symmetry of this sector taken to be $U(N)\times U(N)$,
the effective linear theory contains two dimensionless coupling constants
$\lambda_1$ and $\lambda_2$, which are taken to approach the strong coupling
limit.  The range of couplings considered avoids the Landau pole problem,
and corrections to the linear theory, represented by higher-dimensional
operators, should be relatively small.  Equivalently, a finite ultraviolet
cutoff on the theory can be taken to lie above the momentum range of interest.
When the strong coupling limit is reached (loop expansion parameters equal
unity), higher-dimensional operators are not obviously suppressed and the
scalars described by the linear model are not obviously lighter than other
new physics.  Nevertheless, the strongly first order character of the
transition as the couplings of the linear theory approach unit strength
provides evidence that a theory such as technicolor will indeed give rise
to a strongly first order electroweak phase transition.

   A $1/N$ expansion provides the framework for the computation.
To leading order in $1/N$, all contributions from the $\lambda_1$
interaction can be summed.  With respect to $\lambda_2$, all planar
graphs contribute to leading order in $1/N$, and only a restricted
set of contributions were summed.  We argued that this is adequate to
determine the strongly first order character of the phase transition.
The strength of the transition decreases with $\lambda_2$, consistent
with a second order transition in the $O(2N^2)$ theory.

   The momentum scales in the strongly first order case are such that there
is no reason to expect that a high temperature expansion should be reliable,
or that the phase transition should be governed by an effective
three-dimensional theory.  Our results are nevertheless consistent with the
hints provided by analysis of such theories \cite{PisarskiWilczek,Paterson}.

   This work may be relevant to the problem of electroweak baryogenesis.
More detailed calculations will be necessary to study the dynamics of
departure from thermal equilibrium during the transition, but our strongly
first order transition provides at least a necessary condition for baryon
number generation.  Furthermore, the sphaleron mass in our model is the
same as the minimal standard model formula \cite{msphaleron},
$m_{\rm sphaleron} = (4\pi\, K/g)\langle\sigma_0\rangle$,
where $K\sim 3$ is a function of the scalar couplings.
Since $T_c$ is of order $2 v/N$, a strongly first order phase transition
to a broken phase with $\langle\sigma_0\rangle/v \sim 1$ ensures Boltzmann
suppression of baryon number destruction during and after the phase
transition.

\vskip 1.0cm
\begin{flushleft}{\Large\bf Acknowledgments}\end{flushleft}

   We thank Peter Arnold, Sekhar Chivukula, Andy Cohen, Nick Evans,
Stephen Hsu, Krishna Rajagopal, Yue Shen, John Terning, and Larry Yaffe
for useful discussions and criticism.
T.W.A.\ and S.B.S.\ thank the Aspen Center for Physics for its hospitality
and productive research setting.
This work was supported in part under U.S.\ Department of Energy contract
No.\ DE-AC02-ERU3075.

\end{document}